\documentclass[showpacs,aps,amssymb,floatfix,prd,amsmath,preprintnumbers]{revtex4}
\usepackage{epstopdf}
\usepackage{capt-of}
\usepackage{graphicx}  
\usepackage{dcolumn}   
\usepackage{bm}
\begin{document}
\input epsf.tex
\title{\bf Kaluza-Klein cosmological model in $f(R,T)$ gravity with $\Lambda(T)$}

\author{
P.K. Sahoo\footnote{ Department of Mathematics, Birla Institute of
Technology and Science-Pilani, Hyderabad Campus, Hyderabad 500078,
India,  Email:  sahoomaku@rediffmail.com.}, B.
Mishra\footnote{Department of Mathematics, Birla Institute of
Technology and Science-Pilani, Hyderabad Campus, Hyderabad 500078,
India,  Email:  bivudutta@yahoo.com.}, S.K.
Tripathy\footnote{Department of Physics, Indira Gandhi Institute of
Technology, Sarang, Dhenkanal, Odisha-759146, India Email: tripathy
\_ sunil@rediffmail.com.}}

\affiliation{ }

\begin{abstract}
A class of Kaluza-Klein cosmological models in $f(R,T)$ theory of
gravity have been investigated. In the work, we have considered the
functional $f(R,T)$ to be in the form $f(R,T)=f(R)+f(T)$ with
$f(R)=\lambda R$ and $f(T)=\lambda T$. Such a choice of the
functional $f(R,T)$ leads to an evolving effective cosmological
constant $\Lambda$ which depends on the stress energy tensor. The
source of the matter field is taken to be a perfect cosmic fluid.
The exact solutions of the field equations are obtained by
considering a constant deceleration parameter which leads two
different aspects of the volumetric expansion namely a power law and
an exponential volumetric expansion. Keeping an eye on the
accelerating nature of the universe in the present epoch, the
dynamics  and physical behaviour of the models have been discussed.
From statefinder diagnostic pair we found that the model with
exponential volumetric expansion behaves more like a $\Lambda$CDM
model.

\end{abstract}

\pacs{04.50.kd.}

\keywords{Kaluza-Klein spacetime, $f(R,T)$ gravity, Power law,
Exponential law.}

\maketitle

\section{Introduction}
The late time accelerated expansion of the universe has attracted
much attention in recent years. Direct evidence of cosmic
acceleration comes from high red-shift supernova experiments
\cite{AGR98, AGR04, CLB03}. Some other observations, such as cosmic
microwave background fluctuations \cite{DNS03, DNS07}, baryon
acoustics oscillations \cite{Ein05}, large scale structure
\cite{Teg04} and weak lensing \cite{Jain03} provide a strong support
in favour of  cosmic acceleration. These observations  led to so
many novel ideas and changed the concept regarding the matter-energy
budget of  universe. The accelerated expansion is usually attributed
to a strong negative pressure due to an exotic dark energy form.
Despite a lot of success, Einstein's general relativity (GR) in its
usual form fails to explain the late time acceleration. However, a
cosmological constant (may be time varying) in the source term can
be a good candidate for dark energy. It is now believed that the
universe is mostly dominated by the presence of dark energy
($68.3\%$) and dark matter ($26.8 \%$). The baryonic matter comprises
only of 4.9\%. Of late, modified theories of general relativity seem
attractive in the understanding of the exotic nature of dark energy
and the consequent late time cosmic acceleration. GR can be modified
in two different ways. Either the matter source can be modified (by
considering scalar field contribution or by including an exotic
energy source term in the usual field equations) or by modifying the
underlying geometry. The later method, of late, has gained much
interest where the standard Einstein-Hilbert action is replaced by
an arbitrary function may be of Ricci Scalar $R$ ($f(R)$ gravity),
of scalar torsion $T$ ($f(T)$ gravity), of $G$ ($f(G)$ gravity and
$f(R,G)$ gravity) or of inclusion of some other matter field
Lagrangian along with some geometrical features. For a recent
reviews on modified gravity models one can see \cite{Car04, Sot10,
Noj07} and references therein. Among the various modifications,
$f(R)$ theory of gravity is treated most suitable due to
cosmologically important $f(R)$ models. It has been suggested that
cosmic acceleration can be achieved by replacing the
Einstein-Hilbert action of general relativity with a general
function of Ricci scalar, $f(R)$. A generalization of $f(R)$
modified theories of gravity was proposed \cite{Bert07}, by
including in the theory an explicit coupling of an arbitrary
function of the Ricci scalar $R$ with the matter Lagrangian density
$L_m$. As a result of the coupling the motion of the massive
particles is non geodesic, and an extra force orthogonal to the four
velocity arises.  Nojiri and Odintsov \cite{Noj07} have reviewed
various modified gravity theories that are considered as
gravitational alternatives for dark energy. Many authors
\cite{Mul06, Mul07, Noj11, Cli12} have investigated $f(R)$ gravity
in different context. Shamir \cite{Shamir10} has proposed a
physically viable $f(R)$ gravity model, which show the unification
of early time inflation and late time acceleration.

Recently, Harko et al. \cite{Harko11} proposed a modified theory
of gravity dubbed as $f(R,T)$ gravity. This theory is a generalisation of $f(R)$ theory where the gravitational Lagrangian is taken as an
arbitrary function of the Ricci scalar $R$ and the trace $T$ of
the stress energy tensor. The gravitational
field equations are obtained in metric formalism, which follow from the covariant
divergence of the stress energy tensor. The dependence of trace $T$ may be introduced by exotic imperfect fluids or quantum effects (conformal anomaly). 
The covariant divergence of the stress energy tensor is non zero and as a result the
motion of test particles is non-geodesic. An extra acceleration is always present
due to the coupling between matter and geometry. Recently many authors have investigated different issues in cosmology with late time acceleration in $f(R,T)$ gravity. Adhav \cite{Adhav12} has obtained LRS Bianchi type I
cosmological model in $f(R,T)$ gravity.  In the frame work of
$f(R,T)$ gravity, Reddy et al. \cite{Reddy12} have discussed Bianchi type
III cosmological model  while Reddy et al. \cite{Reddy13}, Reddy and
Shanthikumar \cite{Reddy13a} studied Bianchi type III dark energy model and
some anisotropic cosmological models, respectively. In some recent
works, different aspects of this modified gravity theory have been
investigated by Chaubey and Shukla \cite{Cha13}, Rao and Neelima \cite{Rao13},
Sahoo et al. \cite{Sahoo14}, Mishra and Sahoo \cite{Mishra14}.
Yadav \cite{Yadav13}, Ahmed and Pradhan \cite{Ahmed14} have
studied Bianchi type-V string and perfect fluid cosmological
models respectively by considering $f(R,T)=f_1(R)+f_2(T)$. In their works, Sharif \cite{Sharif12}, Alvarenga \cite{Alv13} and Myrzakulov \cite {Myrz12, Myrz12a} have investigated different aspects of $f(R,T)$ gravity.

In the present work, we have investigated a class of cosmological models described by
Kaluza-Klein space-time for a perfect fluid distribution in the
framework of  $f(R,T)$ gravity. We chose a specific choice of the
functional $f(R,T)=f(R)+f(T)$ with $f(R)=\lambda R$ and $f(T)=\lambda T$. Such a choice of the functional $f(R,T)$ leads to an evolving effective cosmological constant $\Lambda$ which depends on the stress energy tensor. The organisation of this paper is as follows:
The field equations for Kaluza-Klein model in the frame work of $f(R,T)$ gravity are
derived using the particular form  $f(R,T)=f(R)+f(T)$ in Sect. 2 .  We are interested in the dynamics of late time acceleration of the universe, when is it believed that, the deceleration parameter is either slowly varying or almost constant of time. A constant deceleration parameter leads to two different volumetric expansion laws namely: power law  and exponential expansion law.  The dynamics of universe in $f(R,T)$ gravity are studied using power law  and exponential expansion law of volume scale factor in Sect-3 and Sec-4 respectively. The physical behaviour of the models are also discussed. At the end,
summary and conclusions of the work are presented in Sect. 5. In
the present work, we restrict ourselves to the natural unit system
with $G=c=1$, where $G$ is the Newtonian gravitational constant
and $c$ is the speed of light in vacuum.

\section{Field equations for Kaluza-Klein model in $f(R,T)$ gravity}

The action for the modified $f(R,T)$ gravity takes the form
\cite{Harko11,Ahmed14,Sharif12}
\begin{equation}
S=\frac{1}{16\pi}\int {f(R,T)\sqrt{-g}d^{4}x}+\int
{L_{m}\sqrt{-g}d^{4}x},
\end{equation}
where $f(R,T)$ is an arbitrary function of Ricci scalar $(R)$ and the trace  $T=g^{ij}T_{ij}$ of stress-energy tensor $T_{ij}$. $L_m$ is the matter Lagrangian density.

Varying the action (1) with respect to the metric tensor components $g_{ij}$, we get
\begin{equation}
f_R(R,T)R_{ij}-\frac{1}{2} f(R,T) g_{ij} +
(g_{ij}\Box-\nabla_i\nabla_j)f_R(R,T)
=8\pi T_{ij}-f_T(R,T) T_{ij}- f_T(R,T)\theta_{ij},
\end{equation}

where
\begin{equation}
\theta_{ij}= -2T_{ij}+g_{ij}L_{m} - 2g^{\alpha\beta}
\frac{\partial^2 L_m}{\partial g^{ij} \partial g^{\alpha\beta}}.
\end{equation}

Here $f_R(R,T)=\frac{\partial f(R,T)}{\partial R}$,
$f_T(R,T)=\frac{\partial f(R,T)}{\partial T}$.
$\Box\equiv\nabla^i\nabla_i$  is the De Alembert's operator.

The functional $f(R,T)$ depends on the nature of the matter field and hence for different choices of the functional will lead to different models. Since the choice of the functional is arbitrary, in the present work, we consider it in the form $f(R,T)=f(R)+f(T)$ with $f(R)=\lambda R$ and $f(T)=\lambda T$, where $\lambda$ is an arbitrary constant. It is worth to note here that, in his work \cite{Harko11} , Harko et al. considered three different classes of $f(R,T)$ gravity models:

\begin{equation}
f(R,T )= \left\{
\begin{array}{lcl}
    R+2f(T) \\
    f_1(R)+f_2(T) \\
    f_1(R)+f_2(R)f_3(T).
\end{array}
\right.
\end{equation}

The class of models in the present work belong to the second category. The matter source of the universe is usually considered to be perfect fluid for which the energy momentum tensor takes the form $T_{ij}=(\rho+p)u_iu_j-pg_{ij}$. Here  $u^i=(0,0,0,0,1)$ is the velocity vector in comoving coordinates which satisfies the condition $u^iu_i=1$ and
$u^i\nabla_ju_i=0$. $\rho$ and $p$ are energy density and pressure
of the fluid respectively. The matter Lagrangian can be taken
as $L_m=-p$. Now $\theta_{ij}$ in (3)  can be reduced to

\begin{equation}
\theta_{ij}=-2T_{ij}-pg_{ij}
\end{equation}

For the specific choice $f(R,T)=\lambda (R+T)$, the field equation (2) reduces to

\begin{equation}
R_{ij}-\frac{1}{2}Rg_{ij}=\biggl(\frac{8\pi+\lambda}{\lambda}\biggr)T_{ij}+\biggl(p+\frac{1}{2}T\biggr)g_{ij}.
\end{equation}

We choose a small negative value for the arbitrary $\lambda$ to
draw a better analogy with the usual Einstein field equations and we intend to keep this choice of $\lambda$ throughout.

Einstein field equations with cosmological constant term is
usually expressed as
\begin{equation}
R_{ij}-\frac{1}{2}Rg_{ij}=-8\pi T_{ij}+\Lambda g_{ij}.
\end{equation}

A comparison of equations (6) and (7) provides us
\begin{equation}
\Lambda\equiv\Lambda (T)=p+\frac{1}{2}T
\end{equation}
and $8\pi=\frac{8\pi+\lambda}{\lambda}$. In other words, $p+\frac{1}{2}T$ behaves as cosmological constant
and rather than simply being a constant it evolves through the
cosmic expansion. The dependence of the cosmological constant
$\Lambda$ on the trace of the energy momentum tensor $T$ has been
proposed before by Poplawski \cite{Popla06} where cosmological
constant in the gravitational Lagrangian is considered as a function of the
trace energy-momentum tensor, and consequently the model was
denoted $\Lambda(T)$ gravity model. The $\Lambda(T)$ gravity model is consistent with the recent observational data favouring a variable cosmological constant.

We consider a five dimensional Kaluza-Klein metric in the form
\begin{equation}
ds^{2}=dt^{2}-A(t)^{2}(dx^{2}+dy^{2}+dz^{2})-B(t)^{2}d\psi^{2}
\end{equation}
where the fifth coordinate $\psi$ is taken to be space-like.\\
In a co-moving coordinate system the field equations (6), for the
metric (9), can be explicitly written as
\begin{equation}
2\frac {A^{\prime\prime}}{A}+\biggl(\frac
{A^{\prime}}{A}\biggr)^2+ 2\frac {A^{\prime}B^{\prime}}{AB} +
\frac
{B^{\prime\prime}}{B}=\biggl(\frac{8\pi+\lambda}{\lambda}\biggr)p-\Lambda
\end{equation}
\begin{equation}
3\frac {A^{\prime\prime}}{A}+3 \biggl(\frac
{A^{\prime}}{A}\biggr)^2=\biggl(\frac{8\pi+\lambda}{\lambda}\biggr)p-\Lambda
\end{equation}
\begin{equation}
3 \biggl(\frac {A^{\prime}}{A}\biggr)^2 + 3 \frac
{A^{\prime}B^{\prime}}{AB}=-\biggl(\frac{8\pi+\lambda}{\lambda}\biggr)\rho-\Lambda
\end{equation}
where an overhead prime hereafter, denote ordinary differentiation
with respect to cosmic time $"t"$ only. The trace in our model is
given by $T=-4p+\rho$, so that the effective cosmological constant
in equation (8) reduces to
\begin{equation}
\Lambda (T)=\frac{1}{2} (\rho-2p).
\end{equation}
The spatial volume is given by
\begin{equation}
V=a^4=A^3B
\end{equation}
where $a$ is the mean scale factor.\\
Subtracting (10) from (11), we get
\begin{equation}
\frac{d}{dt}\biggl(\frac {A^{\prime}}{A}- \frac
{B^{\prime}}{B}\biggr)+\biggl(\frac {A^{\prime}}{A}- \frac
{B^{\prime}}{B}\biggr)\frac{V^{\prime}}{V}=0
\end{equation}
which on integration yields
\begin{equation}
\frac {A}{B}=c_2 \exp \biggl[c_1 \int \frac {dt}{V}  \biggr]
\end{equation}
where $c_1$ and $c_2$ are integration constants.\\
Using (14), the values of scale factors $A$ and $B$ can be written
explicitly as
\begin{equation}
A=c_2^{1/4}V^{1/4} \exp \biggl[\frac {c_1}{4} \int \frac {dt}{V}
\biggr]
\end{equation}
and
\begin{equation}
B=c_2^{-3/4}V^{1/4} \exp \biggl[\frac {-3c_1}{4} \int \frac
{dt}{V} \biggr]
\end{equation}
The directional Hubble parameters along different directions are
defined as $H_x=\frac {A^{\prime}}{A}=H_y=H_z$ and $
H_{\psi}=\frac {B^{\prime}}{B}$ so that the mean Hubble parameter
becomes $H=\frac{1}{4}(3H_x+H_{\psi})$. In terms of the
directional Hubble parameters, we can express the field
equations(10)-(12) as
\begin{equation}
2H_x^{\prime}+3H_x^2+2H_xH_{\psi}+H_{\psi}^{\prime}+H_{\psi}^2=\alpha
p-\Lambda,
\end{equation}
\begin{equation}
3(H_x^{\prime}+2H_x^2)=\alpha p-\Lambda,
\end{equation}
\begin{equation}
3(H_x^2+H_xH_{\psi})=-\alpha \rho-\Lambda,
\end{equation}
where $\alpha=\frac{8\pi+\lambda}{\lambda}$.

Equations (20), (21) along with eqn.(13) provide us the general
formulations for the physical parameters of the $f(R,T)$ model.
Pressure, energy density and the effective cosmological constant
for the model can be explicitly expressed in terms of the
directional Hubble parameters as
\begin{equation}
p=\frac{3}{\alpha
(2\alpha+3)}\left[(2\alpha+1)H_x^{\prime}+(4\alpha+1)H_x^2-H_xH_{\psi}\right],
\end{equation}
\begin{equation}
\rho=\frac{6}{\alpha
(2\alpha+3)}\left[H_x^{\prime}+(1-\alpha)H_x^2-(\alpha+1)H_xH_{\psi}\right],
\end{equation}
\begin{equation}
\Lambda=-\frac{3}{
2\alpha+3}\left[2H_x^{\prime}+5H_x^2+H_xH_{\psi}\right].
\end{equation}

The ratio of pressure and energy density is usually referred to as
the equation of state and is considered as an important quantity
in the description of dynamics of the universe. The equation of
state parameter can be calculated directly from the expressions of
$p$ and $\rho$ as
\begin{equation}
\omega=\frac{1}{2}\left[\frac{(2\alpha+1)H_x^{\prime}+(4\alpha+1)H_x^2-H_xH_{\psi}}{H_x^{\prime}+(1-\alpha)H_x^2-(\alpha+1)H_xH_{\psi}}\right].
\end{equation}

Deceleration parameter $q=-1-\frac{\dot{H}}{H^2}$ is important quantity in the study of the dynamics of universe. A positive value of the deceleration parameter signifies a decelerating universe whereas its negative value shows that the universe is accelerating. A lot of observational data in recent past has confirmed that the universe is presently undergoing an accelerated phase of expansion and hence it is plausible to think of a negative value of the deceleration parameter. In fact, observations data constrained this parameter to be $q=-0.34\pm 0.05$ in the present time \cite{skumar12}. At late times of cosmic evolution, the deceleration parameter  slowly varies with time or becomes a constant.  It is worth to mention here that, determination of deceleration parameter involves the observation of supernova with high redshift and hence becoming a tough task. Therefore, the time variation  of deceleration parameter still remains an open question. In the present work, we are particularly interested in the dynamics of the late time accelerated universe and assume a constant deceleration parameter eventhough a variable deceleration parameter with a transition from positive value to negative value is considered in many earlier works. A constant deceleration parameter will lead to  models with two different volumetric expansion namely power law expansion and exponential expansion. In the sections to follow, we concentrate on Kaluza-Klein model with a constant deceleration parameter providing power law and exponential volumetric expansion.

\section{Power law expansion}

Equations (10)-(12) are three independent equations in four
unknowns $A, B, \rho, p$ and require some additional conditions to
get viable cosmological models with determinate solutions. Here,
we consider a power law volumetric expansion  \cite{Akarsu10}
\begin{equation}
V=A^3B=t^m,
\end{equation}
which covers all possible expansion histories through out the
evolution of the universe, where $m$ is a positive constant. The
positive nature of the exponent $m$ is in accordance with
observational findings predicting an expanding universe. The power
law expansion of the universe predicts a deceleration parameter
$q=-\frac{a\ddot{a}}{\dot{a}^2}=-1+\frac{4}{m}$. $a$ is the
average scale factor of the universe and is related to the
directional scale factors $A$ and $B$ through $a=(A^3B)^{1/4}$. Keeping an eye on the recent
observational data favouring an accelerating universe, we choose to
go in favour of  a negative deceleration parameter. For the present
model, such a consideration suggests that the parameter $m$ should
be more than 4. The Hubble rate for this model will be
$H=\frac{m}{4t}$, which obviously decreases with the growth of
cosmic time.

The directional scale factors can be obtained by using (26) in
(17) and (18) as
\begin{equation}
A=c_2^{1/4} t^{m/4} \exp \biggl[\frac {c_1}{4(1-m)} t^{1-m}\biggr]
\end{equation}
and
\begin{equation}
B=c_2^{-3/4} t^{m/4} \exp \biggl[\frac {-3c_1}{4(1-m)}
t^{1-m}\biggr]
\end{equation}
Consequently, the directional Hubble parameters in respective
directions are
\begin{equation}
H_x=H_y=H_z=H+\frac{c_1}{4}t^{-m},
\end{equation}
\begin{equation}
H_{\psi}=H-\frac{3c_1}{4}t^{-m}.
\end{equation}

The expansion scalar  $\Theta=4H$ is
\begin{equation}
\Theta=mt^{-1}.
\end{equation}

Pressure $p$, energy density $\rho$ and cosmological constant
$\Lambda$ for the model (9) can now be obtained from (22)-(24)
and (29)-(30) as
\begin{equation}
p(t) =\frac{3}{4\alpha
(2\alpha+3)}\left[(\alpha+1)c_1^2t^{-2m}+m(\alpha m-2\alpha
-1)t^{-2}\right]
\end{equation}
\begin{equation}
\rho(t)=\frac{3}{4\alpha
(2\alpha+3)}\left[(\alpha+2)c_1^2t^{-2m}-m(\alpha
m+2)t^{-2}\right]
\end{equation}
and
\begin{equation}
\Lambda=-\frac{3}{8
(2\alpha+3)}\left[c_1^2t^{-2m}+m(3m-4)t^{-2}\right].
\end{equation}

\subsection{Physical behaviour of the power law model}

The power law expansion of cosmological volume factor  along with
the consideration of an accelerating behaviour suggests that the
exponent $m$ should always be greater than 4. From the expressions
of pressure and energy density in (32) and (33), it is obvious
that the first terms containing the factor $t^{-2m}$ quickly
decrease to extremely small values for large cosmic time $t$ and
the second terms dominate. In other words, at a later epoch,
pressure and energy density behave as $t^{-2}$. That is the
magnitude of pressure and energy density decrease with the growth
of the cosmic time. However, the sign of these quantities depend
upon the choice of the parameters $m$ and $\alpha$. In order to
get the beautiful equation (8), the coupling constant $\lambda$
has to assume a small negative value. To be more accurate, for the
present discussion, we restrict ourselves to the particular value
of $\lambda=-\frac{8\pi}{8\pi +1}$ which gives us $\alpha=-8\pi$.
In figures 1 and 2 we have plotted the time evolution of pressure
and energy density. It is clear from figure 1 that pressure
assumes negative values throughout the evolution of the cosmic
time and it increases from a large negative pressure to small
value at a later epoch. As it is evident from the host of
observational data favouring an accelerated expansion of the
universe, it is believed that a negative pressure is required to
provide an anti gravity effect and to drive the acceleration. On
the other hand, the energy density decreases from some large
positive value in the initial epoch to very small values in later
times.

\begin{figure}[h]
  \hfill
  \begin{minipage}[t]{.45\textwidth}
    \begin{center}
      \includegraphics[width=75mm]{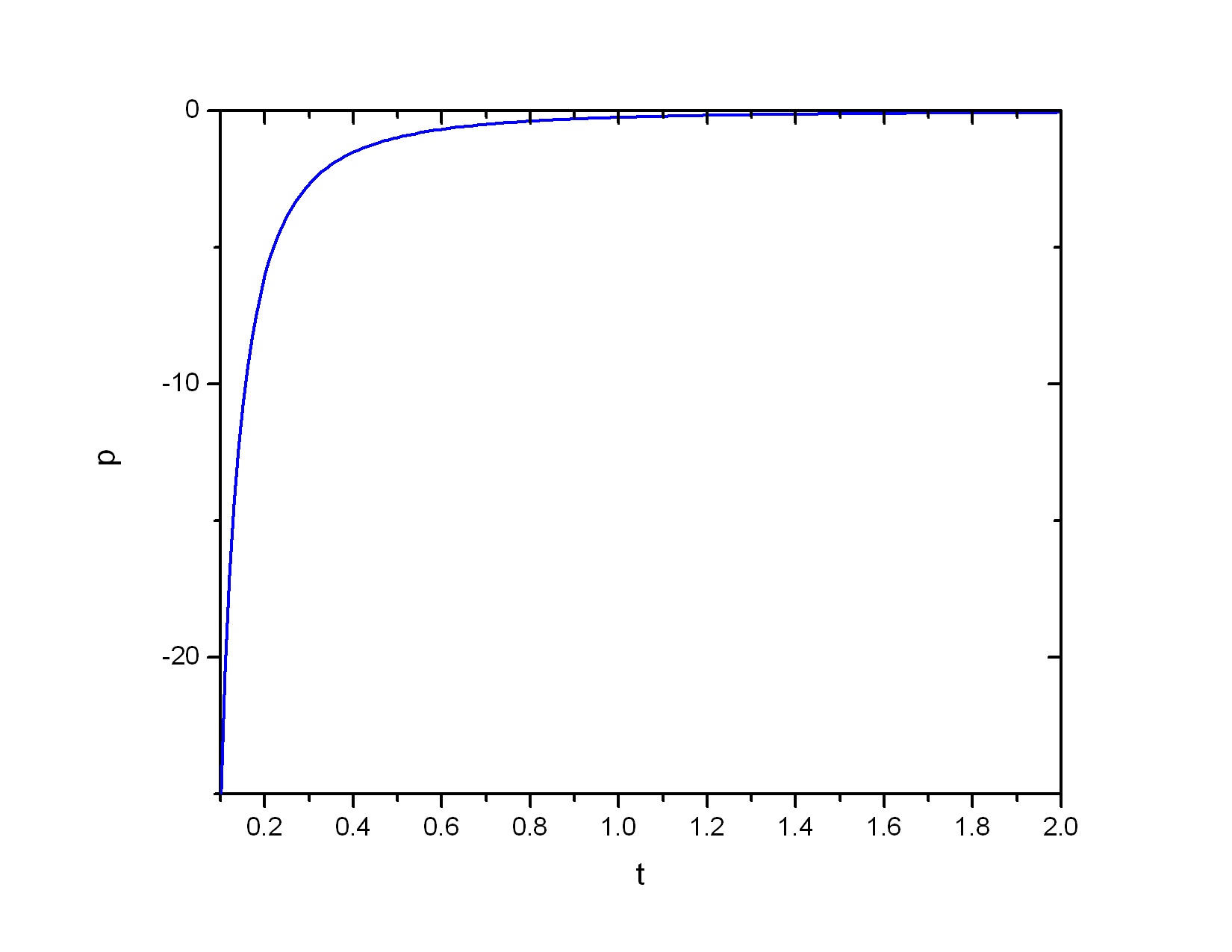}
\caption{Pressure as function of cosmic time}
    \end{center}
  \end{minipage}
  \hfill
  \begin{minipage}[t]{.45\textwidth}
    \begin{center}
      \includegraphics[width=75mm]{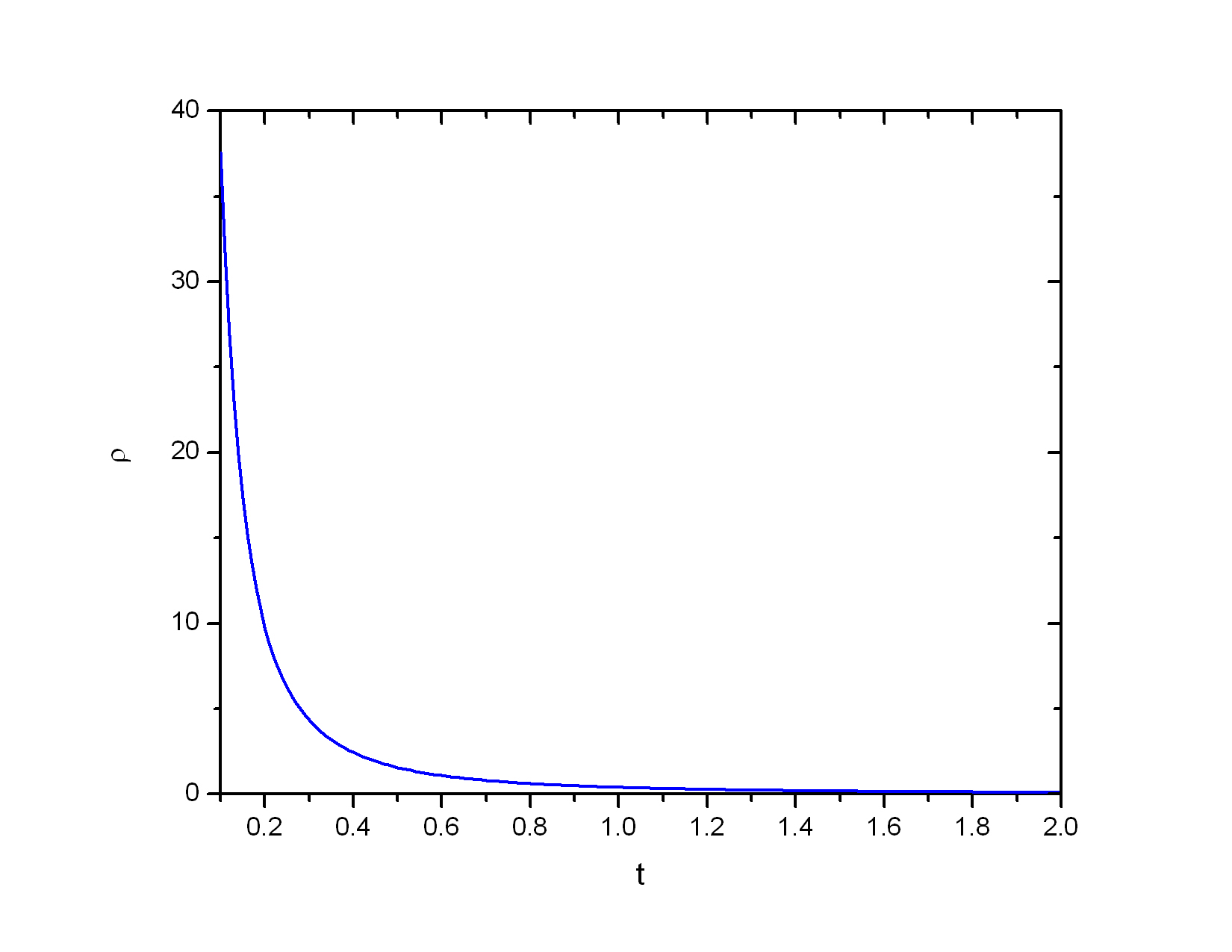}
\caption{Energy density as function of cosmic time}
    \end{center}
  \end{minipage}
  \hfill
 \end{figure}

The cosmic evolution of the cosmological constant is plotted in
figure 3. At a later epoch, the behaviour of cosmological constant
is decided by the second term in equation (34). The cosmological
constant assumes a large value in the initial phase of cosmic
evolution and decrease to small value at later epoch. The
cosmological constant assumes positive value throughout the cosmic
evolution. This sort of behaviour of the cosmological constant is
in accordance to the present accelerated behaviour. Such a behaviour is in consistent with the $\Lambda$CDM model where a small but positive value of the
cosmological constant is required to explain the accelerated
nature of the universe.

\begin{figure}[h!]
\centering
\includegraphics[width=75mm]{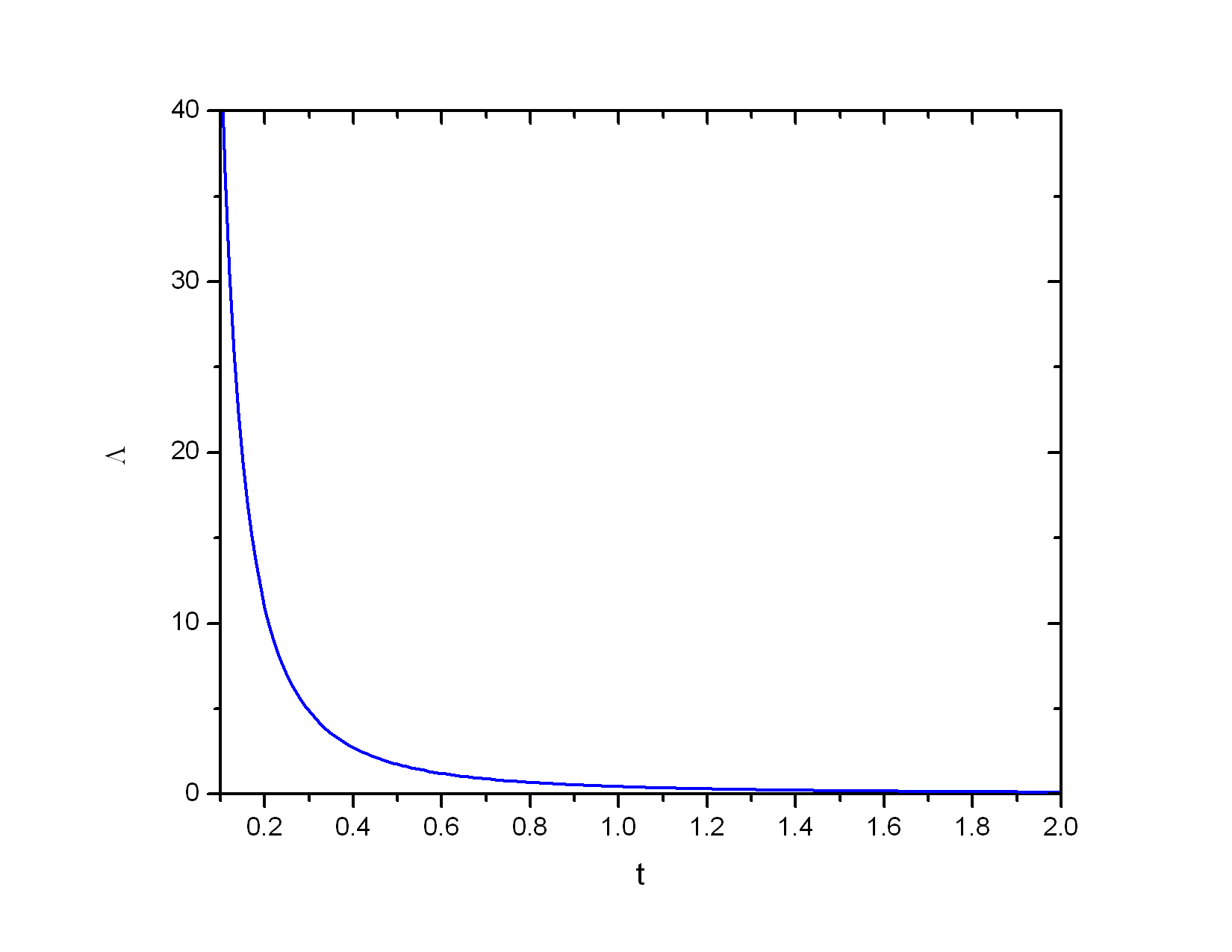}
\caption{Evolution of effective Cosmological constant}
\end{figure}

The equation of state for the power law model becomes

\begin{equation}
\omega=\frac{(\alpha+1)c_1^2t^{-2m}+m(\alpha m-2\alpha
-1)t^{-2}}{(\alpha+2)c_1^2t^{-2m}-m(\alpha m+2)t^{-2}}.
\end{equation}

At a later epoch, when both the first terms in the numerator and
denominator of eqn.(35) containing $t^{-2m}$ subside, the equation
of state behaves as
\begin{equation}
\omega=-1+\frac{2\alpha +3}{\alpha m +2}.
\end{equation}
In figure 4, we have shown the the evolution of the equation of
state parameter for a particular value of the exponent $m$ i.e
$m=5$. The equation of state parameter is negative throughout the
cosmic evolution. The equation of state parameter increases from a
low value at the beginning to an asymptotic value at a later
epoch. It is found from investigation that, with the increase in the value of the exponent $m$, the
asymptotic value decreases. In order to get a clear picture of
evolution of $\omega$, we have plotted it  as a function of
redshift $z$ in figure 5. The redshift $z$ is defined through the
relation $1+z=\frac{1}{a}$, where we have considered the value of
mean scale factor at the present epoch to be 1. In the past,
$\omega$ evolves  from a negative value and increases gradually to
a constant value in future. However, if we restrict ourselves to
the time period encompassing recent past and  future with the
redshift in the range $-0.5\le z \le 2$, $\omega $ remains almost
constant.

\begin{figure}[h]
  \hfill
  \begin{minipage}[t]{.45\textwidth}
    \begin{center}
      \includegraphics[width=75mm]{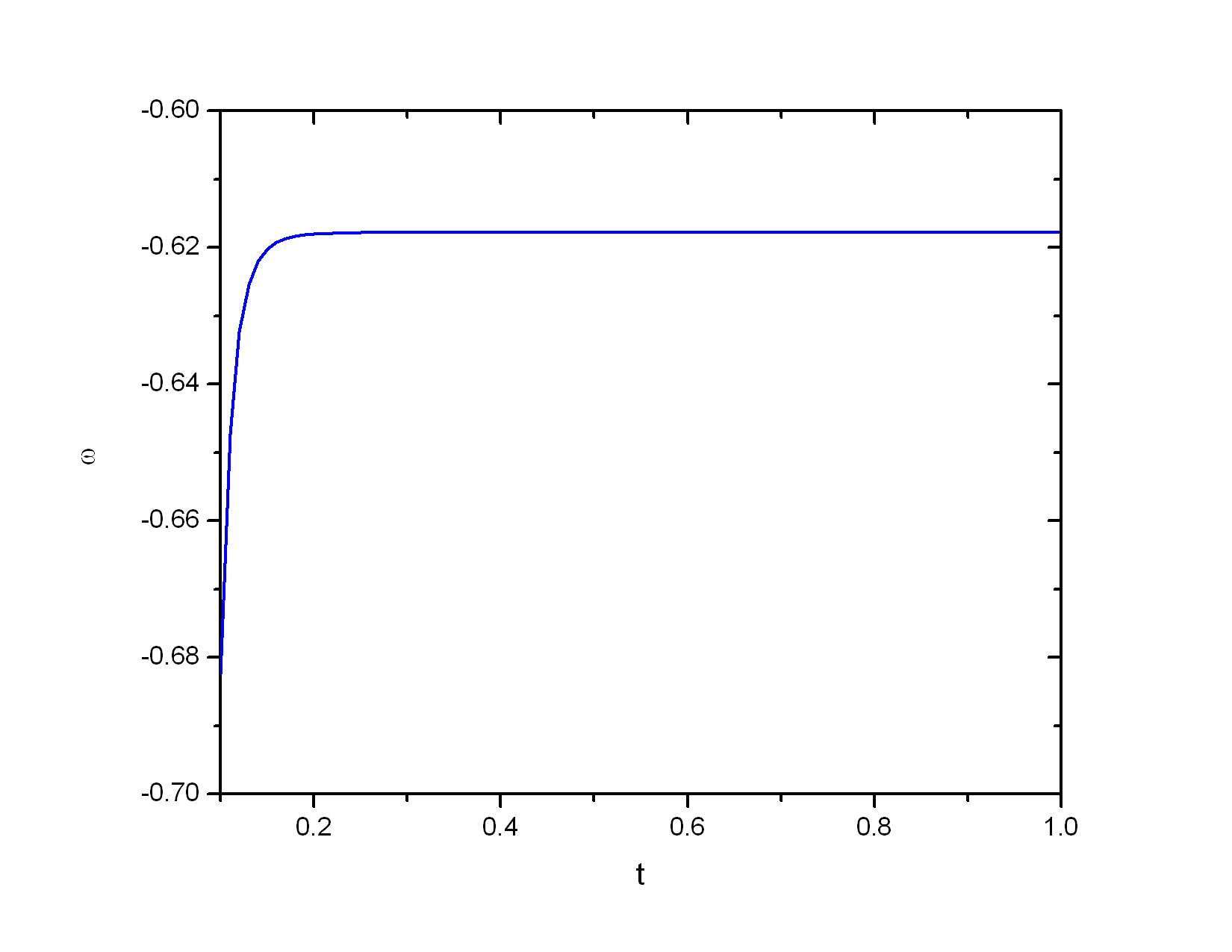}
\caption{Evolution of state parameter as function of cosmic time}
    \end{center}
  \end{minipage}
  \hfill
  \begin{minipage}[t]{.45\textwidth}
    \begin{center}
      \includegraphics[width=75mm]{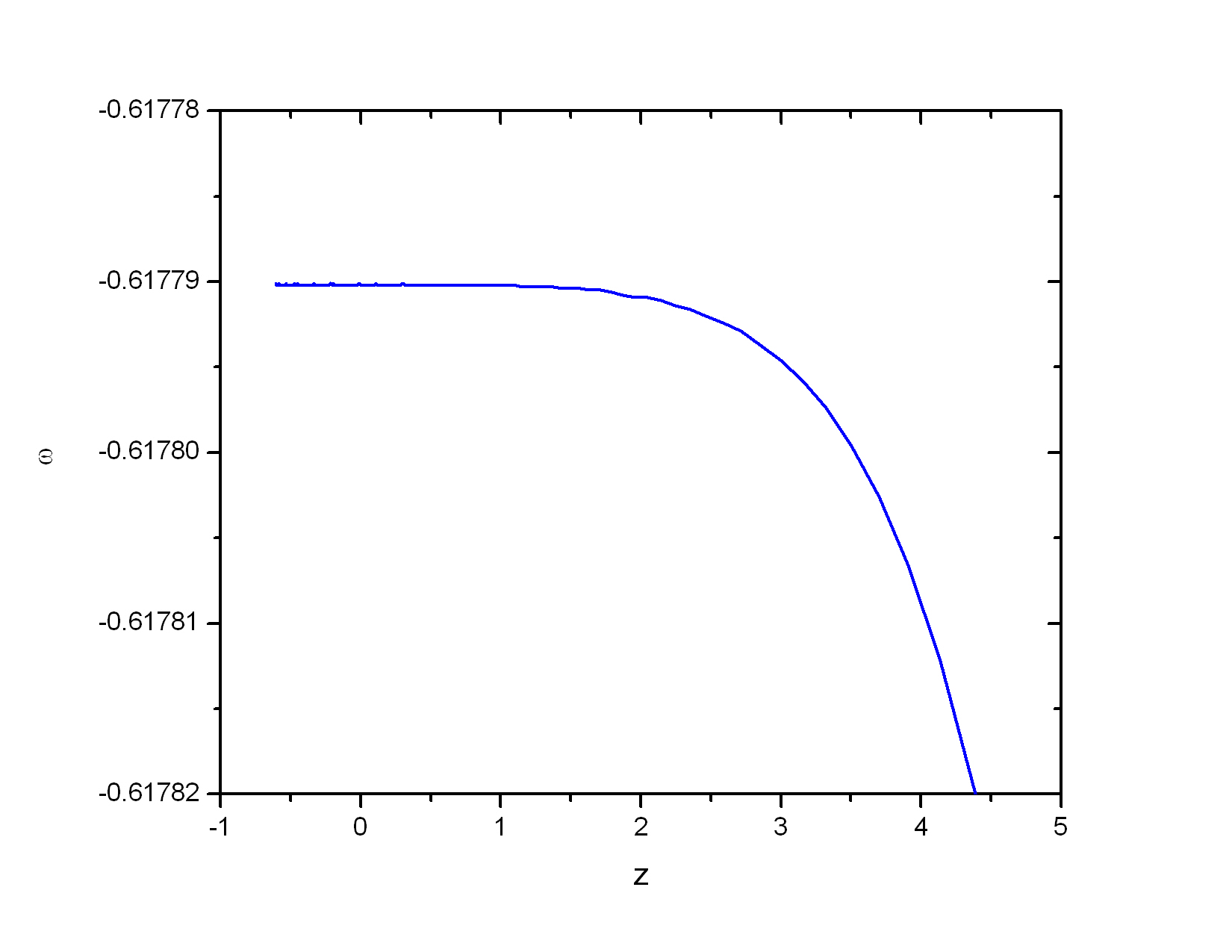}
\caption{Evolution of state parameter as function of redshift}
    \end{center}
  \end{minipage}
  \hfill
 \end{figure}

From the expression of the scalar expansion, it
is evident that, this quantity decreases from some large values
at the beginning to very small values at a later epoch.

In figures 6 and 7, the energy conditions are shown. It is  clear
from the figures that the model with power law volumetric
expansion satisfy the weak energy condition $\rho \ge 0$ and
dominant energy conditions $\rho-p\ge 0 , \rho+p\ge 0$ through out
the cosmic evolution.

\begin{figure}[h]
  \hfill
  \begin{minipage}[t]{.45\textwidth}
    \begin{center}
      \includegraphics[width=75mm]{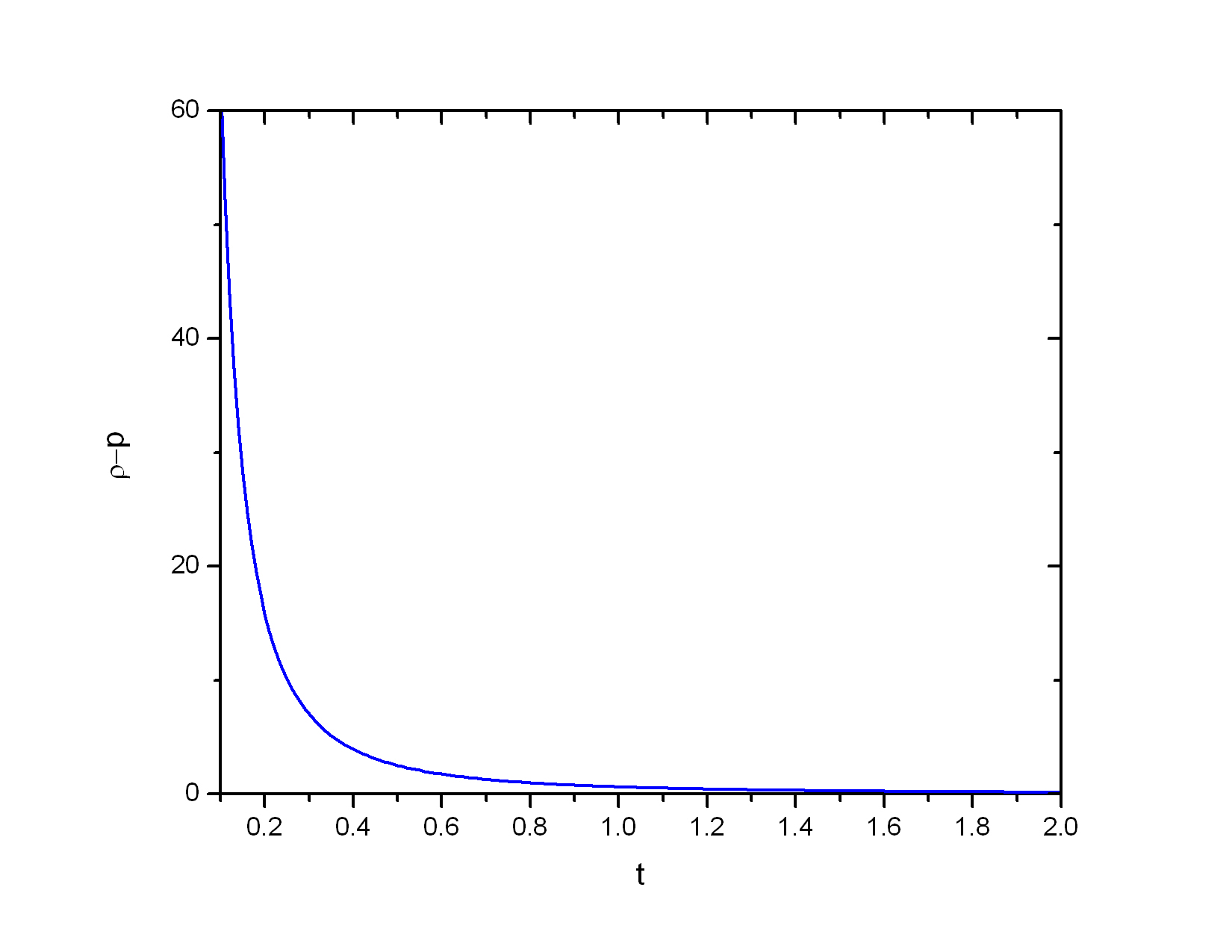}
\caption{Energy condition $\rho-p$ as function of cosmic time}
    \end{center}
  \end{minipage}
  \hfill
  \begin{minipage}[t]{.45\textwidth}
    \begin{center}
      \includegraphics[width=75mm]{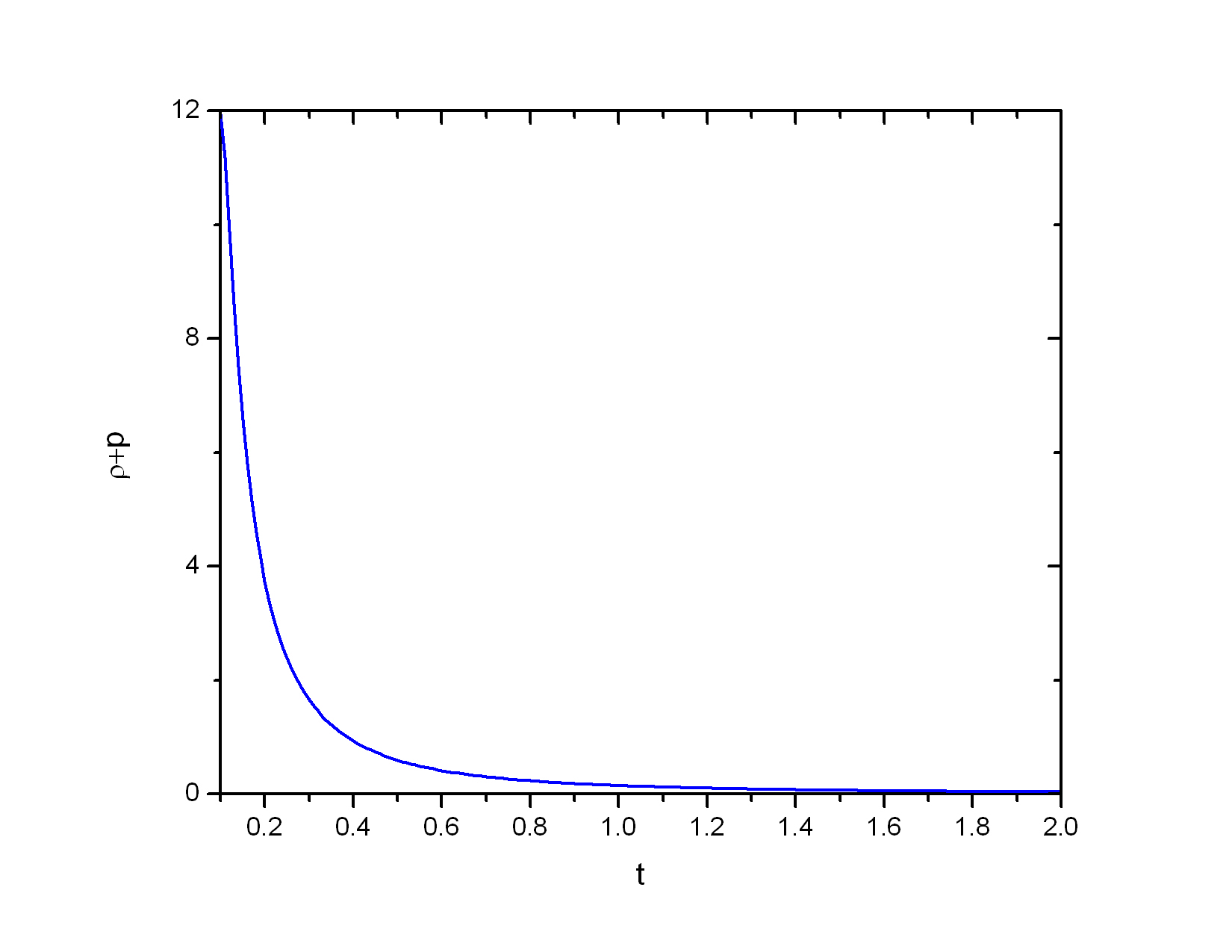}
\caption{Energy condition $\rho+p$ as function of cosmic time}
    \end{center}
  \end{minipage}
  \hfill
 \end{figure}

Statefinder pairs $\{r,s\}$ involve third derivatives of the scale
factor and provide us a better idea about the geometry and becomes a
powerful tool to distinguish between the dark energy models. The
statefinder parameters are defined as
\begin{equation}
r=2q^2+q-\frac{\dot{q}}{H},
\end{equation}
and
\begin{equation}
s=\frac{r-1}{3(q-0.5)}.
\end{equation}
Since the deceleration parameter for this model is
$q=-1+\frac{4}{m}$, $r$ can be expressed as
\begin{equation}
r=1+\frac{32}{m^2}-\frac{12}{m},
\end{equation}
so that $s$ becomes
\begin{equation}
s=\frac{8}{m}.
\end{equation}
The relationship between the two statefinder parameters is
\begin{equation}
r=s\left(\frac{12}{m}+\frac{3m}{8n}-4.5\right).
\end{equation}
The $r-s$ trajectory becomes a straight line with a slope of
$\frac{12}{m}+\frac{3m}{8n}-4.5$. The $\Lambda$CDM model
corresponds to $r=1$ and $s=0$.

\section{Exponential volumetric expansion}

The exponential expansion of volume factor
\begin{equation}
V=e^{4H_0t},
\end{equation}
 usually leads to a de Sitter kind of universe. Here, $H_0$ is the Hubble parameter in the present epoch.
 We have investigated the dynamics of the universe in $f(R,T)$ model with exponential volumetric
 expansion in order to get a clear picture about the dynamics of this model in comparison to the power law model.
 The Hubble parameter is a constant quantity throughout the expansion history of the universe.
 For this choice, the directional expansion rates will be
\begin{equation}
H_x=H_y=H_z=H_0+\frac{c_1}{4}e^{-4H_0t},
\end{equation}
\begin{equation}
H_{\psi}=H_0-3\frac{c_1}{4}e^{-4H_0t}.
\end{equation}
It is interesting to note that, even though the mean Hubble
parameter is a constant quantity, the directional Hubble rates are
time dependent. Similar tendency of Hubble parameter has been
recently investigated \cite{SKT14}. The expansion rate along the
usual axes i.e $x, y$ and $z$ axes decrease from
$H_0+\frac{c_1}{4}$ in the initial epoch to the mean Hubble rate
$H_0$ at a later epoch. On the other hand, the expansion rate
along the extra dimension increase from $H_0-\frac{3c_1}{4}$ to be
equal to the mean Hubble rate at a later epoch. The deceleration
parameter for this model is $q=-1$ and it predicts an accelerated
expansion.

For this model, we get the scale factors as
\begin{equation}
A=c_2^{1/4} \exp \biggl[H_0t-\frac {c_1}{16 H_0}e^{-4H_0t} \biggr],
\end{equation}
and
\begin{equation}
B=c_2^{-3/4} \exp \biggl[H_0t+\frac {3c_1}{16 H_0}e^{-4H_0t}
\biggr].
\end{equation}
Pressure $(p)$, energy density $\rho$ and cosmological constant
$\Lambda$ for the model (9) are calculated to be
\begin{equation}
p(t) =\frac{3}{4\alpha
(2\alpha+3)}\left[(\alpha+1)c_1^2e^{-8H_0t}+16\alpha H_0^2\right],
\end{equation}
\begin{equation}
\rho(t)=\frac{3}{4\alpha
(2\alpha+3)}\left[(\alpha+2)c_1^2e^{-8H_0t}-16\alpha H_0^2\right]
\end{equation}
and
\begin{equation}
\Lambda=\frac{3}{8
(2\alpha+3)}\left[c_1^2e^{-8H_0t}-48 H_0^2\right].
\end{equation}

 The scalar expansion for
this model is obtained as
\begin{equation}
\Theta=4H_0.
\end{equation}

\subsection{Physical behaviour of the exponential law model}

In this model, both the pressure and energy density of the
universe evolve with time. The time dependent terms in the
pressure and energy density are the 1st terms containing the
factor $e^{-8H_0t}$. Since, the value of Hubble rate in the
present epoch is around 70, these terms quickly go to zero and the
constant terms dominate at a later epoch. Therefore in this model,
very quickly a steady state for pressure and energy density is
achieved. Pressure increases from a large negative value in the
beginning of cosmic time and approaches to a constant value
$\frac{12H_0^2}{(2\alpha+3)}$. Similarly, energy density decreases
from a large positive value in beginning of cosmic time to a
constant value $-\frac{12H_0^2}{(2\alpha+3)}$.

The cosmological constant also decreases from large positive value
to a small positive value $-\frac{18H_0^2}{2\alpha+3}$ quickly.
Such a behaviour of the cosmological constant is required to
explain the accelerated nature of the cosmic expansion.

The equation of state parameter for this model is given by

\begin{equation}
\omega=\frac{(\alpha+1)c_1^2e^{-8H_0t}+16\alpha
H_0^2}{(\alpha+2)c_1^2e^{-8H_0t}-16\alpha H_0^2}.
\end{equation}
We have shown the evolution of the equation of state parameter in
figure 8. From the figure it is clear that, $\omega$ lies in the
phantom region. With the growth of time, it asymptotically
increases from a higher negative value to approach $-1$ at late
times. In figure 9, $\omega$ is plotted as a function of redshift
$z$. It may be noted from the figure that, $\omega$ evolves from
certain positive value in the past and at a  critical transition
time it suddenly crosses the phantom divide and becomes negative
and asymptotically it increases to $-1$ in future time. The
statefinder pairs $\{r,s\}$ for this models equal to that of the
$\Lambda$CDM model value $\{1,0\}$.

\begin{figure}[h]
  \hfill
  \begin{minipage}[t]{.45\textwidth}
    \begin{center}
      \includegraphics[width=75mm]{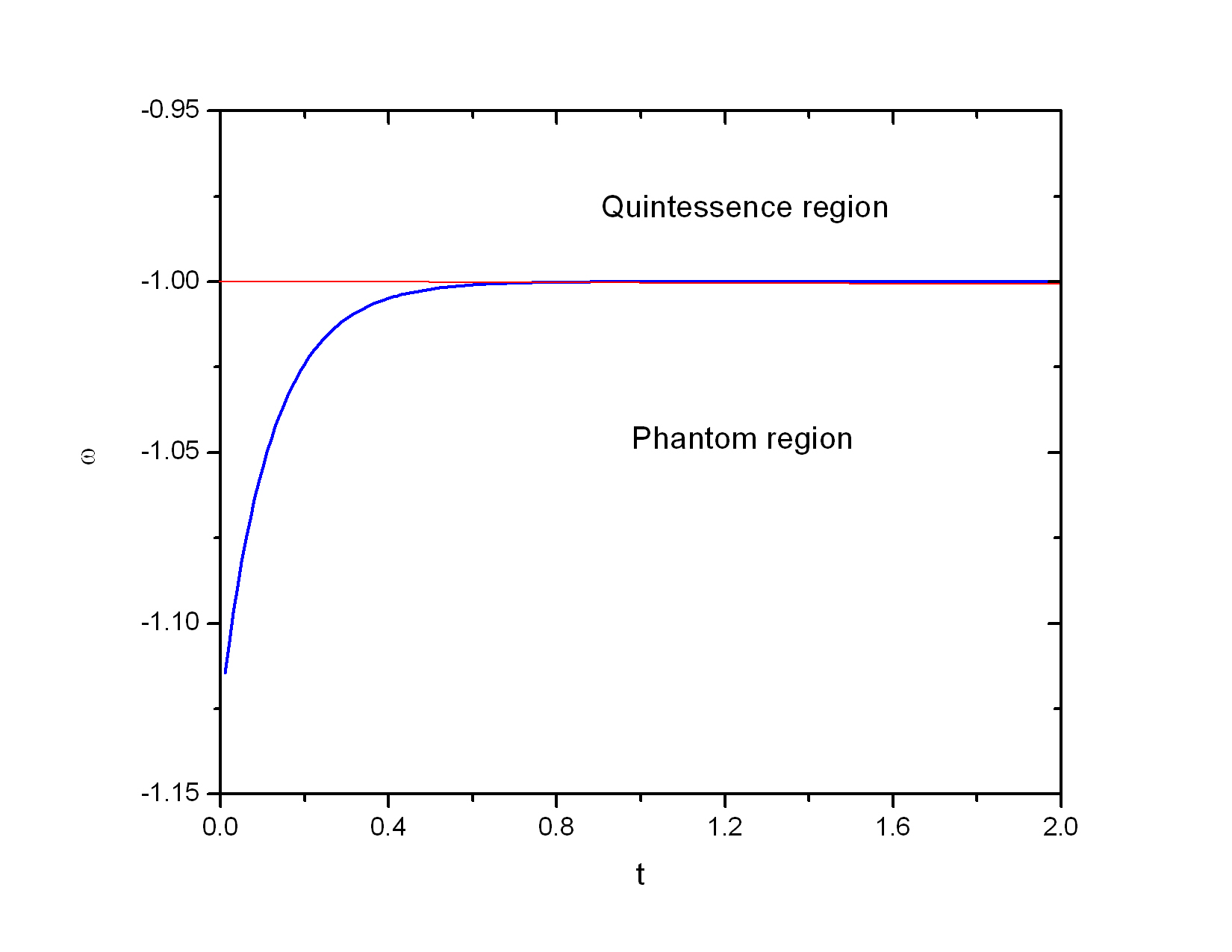}
\caption{Evolution of equation of state parameter for exponential
model. The red line in the graph shows the phantom divide}
    \end{center}
  \end{minipage}
  \hfill
  \begin{minipage}[t]{.45\textwidth}
    \begin{center}
      \includegraphics[width=75mm]{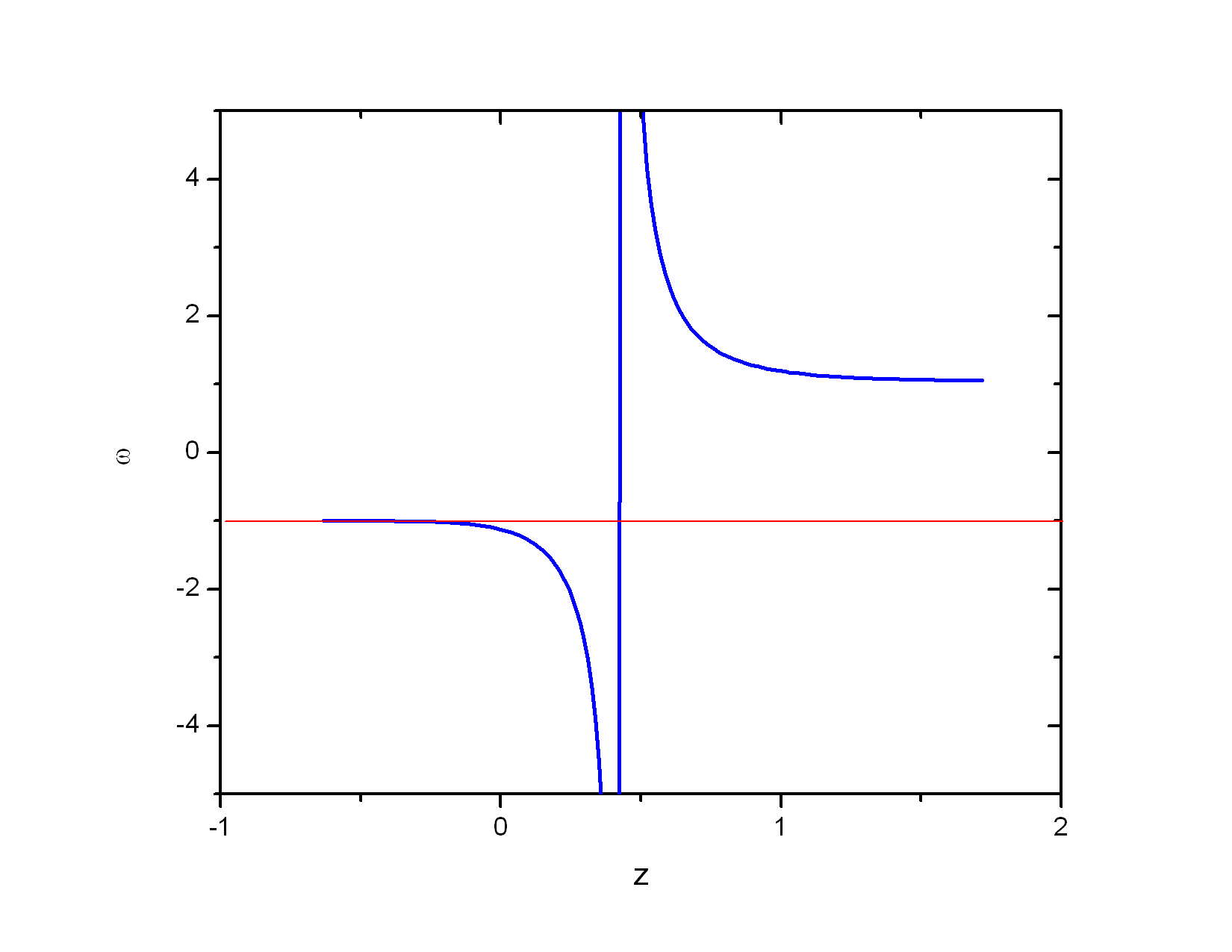}
\caption{Evolution of equation of state parameter as a function of
redshift. The red line in the graph shows the phantom divide}
    \end{center}
  \end{minipage}
  \hfill
 \end{figure}

\section{Conclusion}

The dynamics of a five dimensional Kaluza-Klein metric for a
perfect fluid distribution is  studied in the frame work of a
modified gravity theory, known as $f(R,T)$ gravity model proposed
by Harko et al. \cite{Harko11}. In the $f(R,T)$ theory of gravity,
the gravitational Lagrangian is given by an arbitrary function of
the Ricci Scalar $R$ and of the trace $T$ of the stress energy
tensor. Out of the different plausible choices for the functional
$f(R,T)$, we have used a particular choice $f(R,T)=\lambda(R+T)$,
where $\lambda$ is an arbitrary constant having a small negative
value. We have shown that, for a particular choice of this
parameter, i.e. $\lambda=-\frac{8\pi}{8\pi+1}$,  the field
equations in this model nicely reduce to the usual Relativistic
equations with an effective time dependent cosmological constant.
This effective cosmological constant can then be expressed in
terms of the rest energy density and pressure. In the present
work, we have developed a general  formalism to calculate the
pressure, rest energy density and effective cosmological constant
in terms of the directional Hubble parameters. In  the context of
recent observations predicting an accelerating universe and the
exotic form of dark energy  responsible for cosmic acceleration,
we have studied two different models considering two different
kinds of volumetric expansion namely, power law expansion and
exponential expansion. The physical and geometrical aspects of the
models are also investigated.

In the power law model, it is found that, pressure increases from
a large negative value to an asymptotic value at a later epoch,
whereas in the exponential model, pressure increases from a large
negative value to a constant value. The increment in the later
model is very fast as compared to the power law model. In the
second case, the pressure quickly reaches to a steady state.
Similarly, the energy density in the power law model decreases
from a large positive to very small values at late times. But in
the exponential model, the rest energy density decreases from a
large positive value to a steady state value at a much quicker
rate. In both the models, the effective cosmological constant
decreases from large positive value to small positive value.
However, for exponential expansion of volume factor, the effective
cosmological constant stabilises around a constant value in a late
period of cosmic evolution.

The dynamics of the universe is studied through the equation of
state parameter. In the first case, the equation of state
increases from certain negative value to a steady constant value
which lies above the phantom divide whereas in the second case,
the equation of state parameter asymptotically evolves from below
the phantom divide to the phantom divide and behaves like a
cosmological constant at a late epoch. It is interesting to note
from the discussion that, in the second model, the mean Hubble
parameter is a constant quantity throughout the cosmic evolution
but the directional Hubble parameters are time dependent and
evolve through the expansion history. In the usual dimensions, the
Hubble rate decrease from a large value to the mean constant value
whereas in the extra dimension, the expansion rate exponentially
increase to be equal to the mean Hubble rate at a later time.
From the statefinder diagnosis we found that, the second model
with exponential volumetric expansion, behaves more like a
$\Lambda$CDM model with the statefinder pair $\{r,s\}$ as
$\{1,0\}$ and in the first case the $r-s$ trajectory becomes a
straight line.

\section{Acknowledgement}
PKS would like to thank the Institute of Mathematical Sciences
(IMSc), Chennai, India for providing facility and support during a
visit where part of this work was done. SKT likes to thank
Institute of Physics, Bhubaneswar for providing necessary facility
for accomplishing a part of this work.

\end{document}